\documentclass[manuscript]{aastex}

\def\I{\'\i}
\def\cd{cd$^{-1}$\,}
\def\kms{kms$^{-1}$\,}

\shorttitle{Asteroseismology of $\gamma$ Peg}
\shortauthors{Handler et al.}

\begin{document}

\title{Asteroseismology of hybrid pulsators made possible: \\
simultaneous {\it MOST} space photometry and ground-based spectroscopy
of $\gamma$ Peg\altaffilmark{*}}

\author{G. Handler,\altaffilmark{1} J. M. Matthews,\altaffilmark{2}
J. A. Eaton,\altaffilmark{3} J. Daszy{\'n}ska-Daszkiewicz,\altaffilmark{4} 
R. Kuschnig,\altaffilmark{1} \\ H. Lehmann,\altaffilmark{5} 
E. Rodr\I guez,\altaffilmark{6} A. A. Pamyatnykh,\altaffilmark{1, 7, 8} 
T. Zdravkov,\altaffilmark{7} P. Lenz ,\altaffilmark{1}\\
V. Costa,\altaffilmark{6} D. D\I az-Fraile,\altaffilmark{6}
A. Sota,\altaffilmark{6}
T. Kwiatkowski,\altaffilmark{9} A. Schwarzenberg-Czerny,\altaffilmark{7} 
\\ W. Borczyk,\altaffilmark{9} W. Dimitrov,\altaffilmark{9} 
M. Fagas,\altaffilmark{9} K. Kami\'nski,\altaffilmark{9} 
A. Ro\.zek,\altaffilmark{9} F. van Wyk,\altaffilmark{10}\\
K. R. Pollard,\altaffilmark{11} P. M. Kilmartin,\altaffilmark{11}
W. W. Weiss,\altaffilmark{1} D. B. Guenther,\altaffilmark{12} \\
A. F. J. Moffat,\altaffilmark{13} S. M. Rucinski,\altaffilmark{14} 
D. D. Sasselov,\altaffilmark{15} G. A. H. Walker\,\altaffilmark{16} 
}

\altaffiltext{1}{Institut f\"ur Astronomie, Universit\"at Wien,
T\"urkenschanzstrasse 17, A-1180 Wien, Austria}
\altaffiltext{2}{Department of Physics and Astronomy, University of 
British Columbia, 6224 Agricultural Road, Vancouver, BC V6T 1Z1, Canada}
\altaffiltext{3}{Center of Excellence in Information Systems,
Tennessee State University, Nashville, TN 37209}
\altaffiltext{4}{Instytut Astronomiczny, Uniwersytet Wroc\l awski, ul.
Kopernika 11, 51-622 Wroc\l aw, Poland}
\altaffiltext{5}{Th\"uringer Landessternwarte Tautenburg, 07778
Tautenburg, Germany}
\altaffiltext{6}{Instituto de Astrof\I sica de Andalucia, CSIC, PO Box
3004, 18080 Granada, Spain}
\altaffiltext{7}{Copernicus Astronomical Center, Bartycka 18, 00-716
Warsaw, Poland}
\altaffiltext{8}{Institute of Astronomy, Russian Academy of
Sciences, Pyatnitskaya Str. 48, 109017 Moscow, Russia}
\altaffiltext{9}{Astronomical Observatory, Adam Mickiewicz University,
S\l oneczna 36, 60-286 Poznan, Poland}
\altaffiltext{10}{South African Astronomical Observatory, P. O. Box 9,
Observatory 7935, South Africa}
\altaffiltext{11}{Department of Physics and Astronomy, University of
Canterbury, Private Bag 4800, Christchurch, New Zealand}
\altaffiltext{12}{Department of Astronomy and Physics, St. Mary's
University, Halifax, NS B3H 3C3, Canada}
\altaffiltext{13}{D\'epartment de physique, Universit\'e de Montr\'eal,
C. P. 6128, Succ. Centre-Ville, Montr\'eal, QC H3C 3J7, Canada}
\altaffiltext{14}{Department of Astronomy and Astrophysics, University of
Toronto, 50 St. George Street, Toronto, ON M5S 3H4, Canada}
\altaffiltext{15}{Astronomy Department, Harvard University, 60 Garden
Street, Cambridge, MA 02138}
\altaffiltext{16}{1234 Hewlett Place, Victoria, BC, V8S 4P7, Canada}
\altaffiltext{*}{Based on data from the MOST satellite, a Canadian Space
Agency mission operated by Dynacon, Inc., the Univ.\ of
Toronto Institute of Aerospace Studies, and the Univ.\ of British
Columbia, with assistance from the Univ.\ of Vienna, Austria.}

%% Mark off your abstract in the ``abstract'' environment. In the manuscript
%% style, abstract will output a Received/Accepted line after the
%% title and affiliation information. No date will appear since the author
%% does not have this information. The dates will be filled in by the
%% editorial office after submission.

\begin{abstract} We have acquired simultaneous high-precision space
photometry and radial velocities of the bright hybrid $\beta$~Cep/SPB
pulsator $\gamma$~Peg. Frequency analyses reveal the presence of six g
modes of high radial order together with eight low-order $\beta$~Cep
oscillations in both data sets. Mode identification shows that all
pulsations have spherical degrees $\ell=0-2$. An 8.5~M$_\sun$ model
reproduces the observed pulsation frequencies; all theoretically predicted
modes are detected. We suggest, contrary to previous authors, that
$\gamma$~Peg is a single star; the claimed orbital variations are due to
g-mode pulsation. $\gamma$~Peg is the first hybrid pulsator for which a
sufficiently large number of high-order g modes and low order p and mixed
modes have been detected and identified to be usable for in-depth seismic
modeling. \end{abstract}

%% Keywords should appear after the \end{abstract} command. The uncommented
%% example has been keyed in ApJ style. See the instructions to authors
%% for the journal to which you are submitting your paper to determine
%% what keyword punctuation is appropriate.


\section{Introduction}

The bright ($V=2.8$) B2 IV star \object{$\gamma$~Peg} was recognized as a
pulsating variable of the $\beta$~Cep class more than 50 years ago
\citep{nam53}, and it was believed to be singly periodic until recently.
\citet{cha06} studied the star spectroscopically and demonstrated its
multiperiodicity. These authors also examined the claim that $\gamma$~Peg
is a spectroscopic binary and deduced an eccentric ($e=0.62$) 370.5-day
orbit. This orbital solution was disputed by \citet{bp07} who suggested
that the orbital eccentricity was spurious and caused by outbursts similar
to those of Be stars, and favored an orbital period near 6.8~d, in
accordance with the original suggestion by \citet{har79}.

The multiperiodic oscillations of \object{$\gamma$~Peg} are highly
interesting because they are caused by two different sets of pulsation
modes: two frequencies detected by \citet{cha06} correspond to
low-order pressure (p) and gravity (g) modes typical for $\beta$~Cep
stars, but the other two are high-order g~modes as excited in the Slowly
Pulsating B (SPB) stars. Indeed, \object{$\gamma$~Peg} is located in the
overlap region of both types of variables in the HR diagram (see
\citet{h09}). The frequencies of the two sets of modes are sensitive to
the physical conditions in different parts of the stellar interior. Hybrid
oscillators therefore offer the possibility to obtain a more complete
picture of the physics inside a star using asteroseismology where
pulsations act as seismic waves (see, e.g., \citet{dp08} for case
studies).

However, the possible binarity of \object{$\gamma$~Peg} imposes
difficulties. The detection of low-frequency oscillations can be
compromised by an inaccurate orbital solution. Photometry does not suffer
from this problem (provided the light-time effect is negligible).
Consequently, \citet{h09} carried out a multicolor time-series photometric
study of \object{$\gamma$~Peg}, detected four SPB-type pulsation modes,
and confirmed the two modes of $\beta$~Cep type. One of the latter was
identified as radial; it would be either the fundamental mode or the first
overtone, immediately constraining the mean stellar density. The enormous
asteroseismic promise of \object{$\gamma$~Peg} motivated us to perform a
high-precision photometric and radial velocity study of its pulsations, in
the hope to detect a sufficient number of p and g modes to sound its
interior structure.

%\clearpage

\section{Observations}

\object{$\gamma$~Peg} was observed with the MOST satellite
\citep{most03} from Sep 16 to Oct 16, 2008. Due to the brightness of the
target, Fabry imaging mode was used, and the data reduction method
developed by \citet{RKF06} to minimize the effects of stray light was
employed. There were over 55000 data points at a cadence of 30~s in the
original data set. These were summed into 4018 0.005-d bins having an rms
scatter of 1~mmag per point and an effective Nyquist frequency of 
$\sim$92\,\cd.

A simultaneous ground-based spectroscopic multisite campaign was organized
for pulsational mode identification. The majority of the data originated
from the Automatic Spectroscopic Telescope of Tennessee State University
\citep[TSU-AST,][]{eat07}, and consisted of 1660 spectra (4900--7100 \AA)  
taken over a span of ten weeks (Sep 3 - Nov 15, 2008) with an effective
Nyquist frequency of $\sim$68\,\cd. We have reduced these spectra and
extracted radial velocities from 34 lines with the techniques described by
\citet{eat07}. The external error of these radial velocities is about
0.2\,\kms. This high precision results from $\gamma$~Peg having a very
sharp lined spectrum for a hot star; \citet{T06} listed $v \sin i =
0$\,\kms! Concerning the star's metallicity, \citet{M06} derived 
$Z=0.0091\pm0.0021$ from optical and \citet{JDDN05} determined 
$[m/H]=-0.04\pm0.08$ from ultraviolet spectra.

Additional spectroscopy was carried out at three observatories, and
ground-based time-resolved multicolor photometry was acquired at three
more sites. However, the present paper only reports the initial results
from the MOST space photometry and TSU-AST radial velocities.

\section{Frequency analysis}

The heliocentrically corrected data were searched for periodicities using
the program {\tt Period04} \citep{LB05}. Amplitude spectra were computed,
compared with the spectral window functions, and the frequencies of the
intrinsic and statistically significant peaks in the Fourier spectra were
determined. Multifrequency fits with all detected signals were calculated
step by step, the corresponding frequencies, amplitudes and phases were
optimized and subtracted from the data before computing residual amplitude
spectra, which were then examined in the same way.

\begin{figure}
\centering
\includegraphics[angle=0,scale=.75,clip]{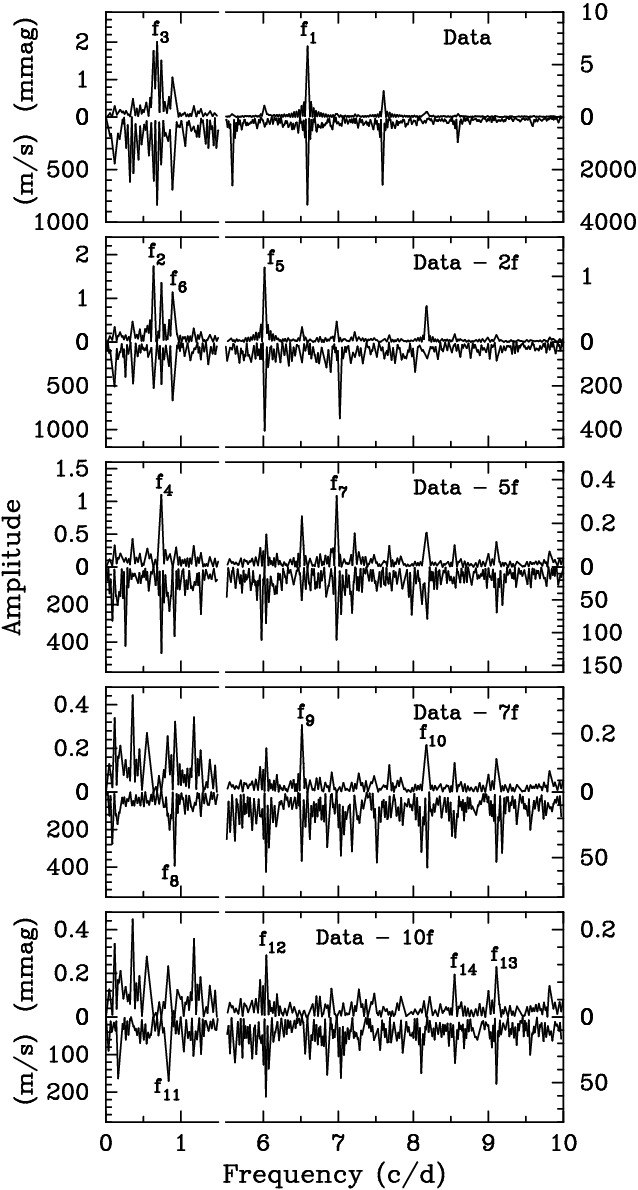}
\caption{Amplitude spectra of our $\gamma$ Peg data with consecutive
prewhitening. The photometric data are compared with the radial velocity
measurements (reverted graphs). The frequency regions $1.5-5.5$ and
$>10$\,\cd are not shown as they contain no intrinsic signals. The
frequencies $f_1 - f_6$ have been assigned for consistency with 
\citet{h09}. The apparent 1\,\cd sidelobe of $f_1$ in the MOST data is an 
alias peak originating from the orbital frequency.\label{fig1}} 
\end{figure}

This analysis was performed for the MOST and radial velocity data
independently. We conservatively only accepted signals that exceeded an
amplitude signal-to-noise ratio of five in at least one of the data sets
and that were prominent in the other. Some steps of this procedure are
shown in Fig.\ \ref{fig1}. Fourteen independent signals were detected; the
agreement between the photometric and radial velocity measurements is
remarkable. The residual amplitude spectrum after this solution was
featureless for the radial velocities. Some peaks in the residual MOST
data remained, but to err on the side of caution we did not consider them
to be intrinsic to $\gamma$~Peg.

%% HERE GOES Fig. 1

The frequencies of all these signals are consistent within the errors
between the two data sets. We have determined weighted mean values of
those frequencies by computing the formal uncertainties \citep{MO99} in
the individual data sets and then applying their inverse squared as the
weight. With the resulting frequencies fixed, we have re-calculated the
amplitudes, phases and amplitude signal-to-noise ratios and list the
results in Table~\ref{tbl-1}.

%% HERE GOES TABLE 1

All but the two closest frequencies are resolved within our data set.
Their difference $f_{12} - f_5 = 0.0112$~\cd is 82\% of the time
resolution of the radial velocity data. Since a significant peak remains
after prewhitening $f_5$ we accept $f_{12}$, but caution that the
parameters of these two variations in Table~\ref{tbl-1} may have
systematic errors.

\section{Discussion}

We start by examining previous claims that $\gamma$~Peg is a spectroscopic
binary. Our radial velocities were acquired in a period of time where the
orbital solution by \citet{cha06} predicts a change in radial velocity
of about 20\,\kms. Our nightly mean radial velocities are constant within
$\pm 0.3$\,\kms. \citet{bp07} claimed an orbit with a 6.816\,d period and
about 0.8\,\kms radial velocity amplitude. Again, this is inconsistent
with our data. Instead, the 1\,\cd alias of this "orbital" period lies
well within the domain of the g-mode frequencies
($1-1/6.816$\,d=0.8533\,\cd); the strongest g-mode pulsations have radial
velocity amplitudes of 0.7\,\kms. We conclude that $\gamma$~Peg is not a
6.8-d spectroscopic binary. The cause of the sporadic radial velocity
changes remains to be understood. The possibility of Be-star outbursts
\citep{bp07} seems remote given that $\gamma$~Peg likely rotates
intrinsically slowly, unlike the Be stars (e.g., \citet{P96}).

\begin{figure}
\centering
\includegraphics[angle=0,scale=.95,clip]{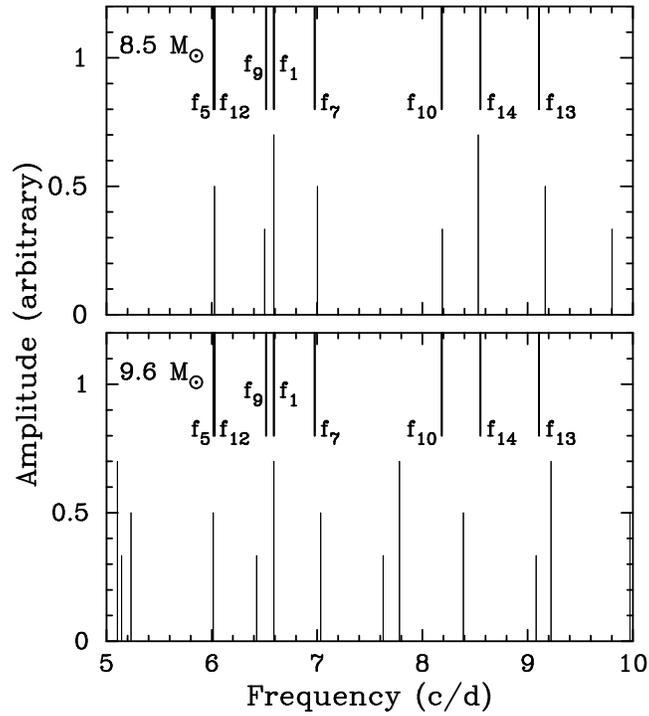}
\caption{Comparison of theoretically predicted pulsation frequencies of
two models fitting $f_1$ and $f_5$. The model frequencies are plotted at
the bottom of the panels, with amplitudes of 0.7 (radial modes), 0.5
(dipole modes), and 0.33 (quadrupole modes). The observed frequencies
protrude from the top of the panels.\label{fig2}} 
\end{figure}

According to the position of $\gamma$~Peg in the HR diagram determined by
\citet{h09}, all signals with frequencies below 1\,\cd are due to
high-order g~modes; the remaining variations are caused by low-order p and
mixed modes. \citet{h09} identified $f_1$ as a radial mode and argued that
$f_5$ is a dipole mode; these are the only two modes with reasonably
secure identifications. From comparison with pulsation models this author
concluded that models in just two small domains of parameter space in the
HR diagram are consistent with these mode identifications and with the
frequency values. In case $f_1$ is the radial fundamental mode, models
around 8.5~M$_{\sun}$ match the observations. If $f_1$ corresponded to the
first radial overtone, models with masses around 9.6~M$_{\sun}$ reproduce
$f_1$ and $f_5$. All models in the present paper and by \citet{h09} were
computed with the the Warsaw-New Jersey stellar evolution and pulsation
code (e.g., see \citet{P98} for a description), using OP opacities, the
\citet{a04} element mixture, and providing linear nonadiabatic model
frequencies. No rotation or convective core overshooting was included in
the models for simplicity of this preliminary model fitting. Figure
\ref{fig2} compares the additional $\ell=0-2$ $\beta$~Cep-type pulsation
frequencies predicted by the two models with our new observations.

%% HERE GOES Fig. 2

All newly detected pulsation frequencies are explained by the 8.5
M$_{\sun}$ model (upper panel of Fig. \ref{fig2}). At first sight, there
are two small inconsistencies: the close doublet $f_5/f_{12}$ only has one
theoretical counterpart, and $f_{13}$ is not well matched by the
theoretical value. The doublet can be explained by possible rotational
splitting at $v_{rot} \approx 3$\,\kms, consistent with the very low $v
\sin i$ of the star. The mismatch for $f_{13}$ may be more of a problem,
but it would not be new: in their analysis of the pulsation spectrum of
\object{$\nu$~Eri}, \citet{PHD04} noted the same problem for the
highest-frequency p~mode ($\ell=1$, p$_2$) observed. Interestingly, in the
present model $f_{13}$ would also correspond to the $\ell=1$, p$_2$ mode
and even the size of the frequency mismatch is similar. It can be 
suspected that this mismatch originates from inadequate physics in the 
models that can be improved through asteroseismology.

% However, it must be added that this model also
% possesses an $\ell=4$ mode very close to $f_{13}$.

The 9.6 M$_{\sun}$ model (lower panel of Fig. \ref{fig2}) results in
poorer agreement between the observed and theoretical frequencies,
although the number of theoretically predicted modes in the domain of
excited frequencies is larger. In particular, only one theoretical
$\ell=1$ mode is available to match $f_{10}$ and $f_{14}$. One would need
to invoke rather fast rotation to explain both modes, but in this case
$\gamma$~Peg would be viewed close to pole-on, which would cause heavy
geometrical cancellation of just these pulsations. We conclude that, in
all likelihood, $\gamma$~Peg is a $\sim$ 8.5 M$_{\sun}$ star oscillating
with a dominant radial fundamental mode. In this case, all the
theoretically predicted $\ell = 0 - 2$ $\beta$~Cep-type pulsation modes
are observed.

\begin{figure}
\centering
\includegraphics[angle=0,scale=.95,clip]{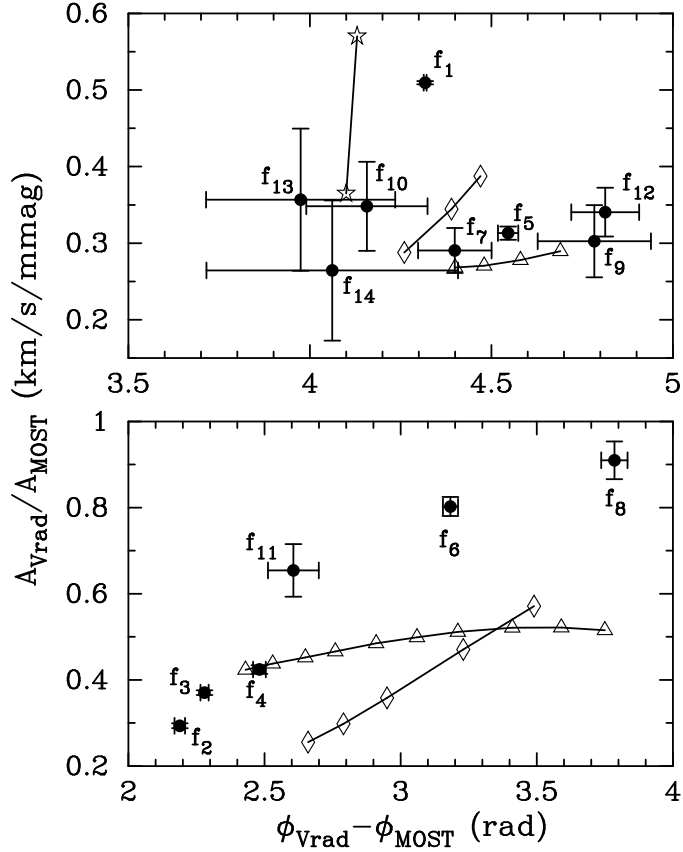}
\caption{Observed (full circles with error bars) and theoretically
predicted radial-velocity to photometric amplitude ratios and phase shifts
for the fourteen signals detected. The upper panel shows the $\beta$~Cep  
pulsations and the lower panel the SPB oscillation modes. The
star symbols show the theoretical locations of radial modes in this
diagram, the diamonds represent dipole modes and the triangles
stand for quadrupole modes. Only model modes with frequencies in the
observed domains are plotted. \label{fig3}} 
\end{figure}

The amplitudes and phases of the oscillations can be used to type the
pulsation modes. We have computed theoretical radial velocity to light
amplitude ratios and phase differences with for the {\it MOST} bandpass.
The 8.5 M$_{\sun}$ model was used together with static atmospheres
\citep{K04} with a metallicity parameter $[m/H]=0.0$ and a microturbulence
velocity $\xi_t=2$ km/s. The comparison of the theoretical and observed
amplitude ratios and phase shifts is shown in Fig.\ \ref{fig3}; modes with
$0 \leq \ell \leq 4$ were considered.

%% HERE GOES Fig. 3

The measured amplitude ratios and phase shifts are consistent with the
interpretation that all oscillation frequencies are caused by modes with
$\ell \leq 2$; theoretical results for modes with higher $\ell$ are off
scale in Fig.\ \ref{fig3}. This corroborates the identification obtained
in Fig.\ \ref{fig2}.

Together with our ground-based multicolor photometry, our radial
velocities can be used to constrain nonadiabatic pulsation theory, as
empirical determinations of the $f$ parameter used to describe the
bolometric flux amplitude depending on the surface displacement can be
made. In turn, this can be used to choose the most suitable opacities for
model calculations \citep{DPP05}.

\begin{figure}
\centering
\includegraphics[angle=0,scale=.95,clip]{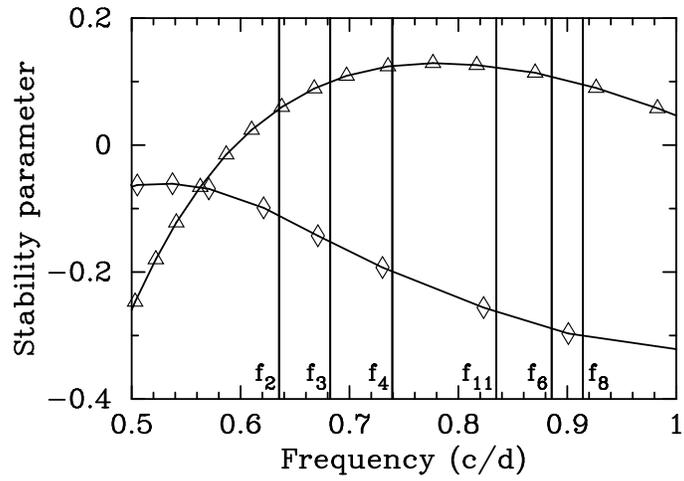}
\caption{Comparison of theoretically predicted g-mode pulsation
frequencies for a 8.5~M$_{\sun}$ model of $\gamma$~Peg with the observed
frequencies. As ordinate the stability parameter $\eta$ is used; if it is 
larger than zero, the corresponding mode is excited in the model. Diamonds 
represent dipole modes and triangles stand for quadrupole modes. The 
vertical lines are drawn at and identified with the observed 
frequencies.\label{fig4}}
\end{figure}

Figure \ref{fig4} shows a comparison of theoretically predicted g-mode
frequencies of our 8.5~M$_{\sun}$ model with the observations. The
frequency separations of the g~modes are well explained by this model:
five of the six observed modes may form a sequence of consecutive radial
overtones of $\ell=1$ modes. The sixth mode would then be $\ell=2$. The
$\ell=1$ modes are stable; a final seismic model must explain their
excitation.

%% HERE GOES Fig. 4

We conclude that $\gamma$~Peg presents sufficient information to carry out
detailed seismic modeling of a hybrid pulsator for the first time. The
eight $\beta$~Cephei pulsation frequencies restrict the possible models
and their parameters considerably. The observed amplitude ratios and phase
shifts between the radial velocities and photometric data provide clues
towards the opacities to be used. Theoretical models constrained by such a
large set of observables must also reproduce the high-order g~modes and
the excited frequency domains. The present study is a demonstration of the
value of combining space photometry with ground-based spectroscopy of
bright stars.

%% Observe the use of the LaTeX \label
%% command after the \subsection to give a symbolic KEY to the
%% subsection for cross-referencing in a \ref command.
\acknowledgments

This work has been supported by the Austrian Fonds zur F\"orderung der
wissenschaftlichen Forschung under grant P20526-N16. AFJM is grateful for
financial support from NSERC (Canada) and FQRNT (Quebec). We thank Frank
Fekel of Tennessee State University for providing the list of photospheric
lines we used for extracting velocities from the TSU spectra. We are
grateful to Ramotholo Sefako for organizing the SAAO observations.

{\it Facilities:} \facility{TSU:AST} \facility{MOST} \facility{TLS}

\clearpage

\clearpage

\clearpage

\clearpage

%% Tables should be submitted one per page, so put a \clearpage before
%% each one.

\clearpage

\begin{deluxetable}{lcccccc}
%\centering
%\tabletypesize{\scriptsize}
%\rotate
\tablecaption{Multifrequency solution for our photometric and radial velocity
data.\label{tbl-1}} 
\tablewidth{0pt}
\tablehead{
\colhead{ID} & \colhead{Frequency} & \colhead{Amplitude} & \colhead{Amplitude} &
\colhead{$(S/N)_{phot}$} & \colhead{$(S/N)_{RV}$} &
\colhead{$\phi_{Vrad} - \phi_{MOST}$} \\
\colhead{} & \colhead{(\cd)} & \colhead{(mmag)} & \colhead{(\kms)} & \colhead{} &
\colhead{} & \colhead{(rad)}}
\startdata
$f_1$ & 6.58974 $\pm$ 0.00002 & 6.59 & 3.359 & 300.9 & 327.9 & 4.318 $\pm$ 0.004 \\ 
$f_2$ & 0.63551 $\pm$ 0.00010 & 1.70 & 0.500 &  21.3 &  17.5 & 2.189 $\pm$ 0.019 \\ 
$f_3$ & 0.68241 $\pm$ 0.00007 & 1.99 & 0.736 &  24.8 &  26.0 & 2.280 $\pm$ 0.015 \\ 
$f_4$ & 0.73940 $\pm$ 0.00010 & 1.23 & 0.522 &  15.4 &  18.6 & 2.481 $\pm$ 0.023 \\ 
$f_5$ & 6.01616 $\pm$ 0.00014 & 1.14 & 0.358 &  45.0 &  34.4 &  4.55 $\pm$ 0.03 \\ 
$f_6$ & 0.88550 $\pm$ 0.00007 & 0.90 & 0.723 &  11.3 &  25.9 &  3.18 $\pm$ 0.03 \\
$f_7$ & 6.9776 $\pm$ 0.0005 & 0.33 & 0.095 &    16.3 &   9.5 &  4.40 $\pm$ 0.10 \\ 
$f_8$ & 0.91442 $\pm$ 0.00011 & 0.51 & 0.464 &   6.4 &  16.6 &  3.79 $\pm$ 0.05 \\ 
$f_9$ & 6.5150 $\pm$ 0.0008 & 0.21 & 0.063 &     9.4 &   6.3 &  4.78 $\pm$ 0.15 \\ 
$f_{10}$ & 8.1861 $\pm$ 0.0008 & 0.18 & 0.064 &  9.2 &   6.7 &  4.16 $\pm$ 0.17 \\ 
$f_{11}$ & 0.8352 $\pm$ 0.0003 & 0.27 & 0.180 &  3.4 &   6.3 &  2.61 $\pm$ 0.09 \\ 
$f_{12}$ & 6.0273 $\pm$ 0.0005: & 0.33 & 0.112 & 12.4 & 10.3 &  4.81 $\pm$ 0.10 \\ 
$f_{13}$ & 9.1092 $\pm$ 0.0012 & 0.12 & 0.041 &  5.9 &   4.5 &  3.97 $\pm$ 0.26 \\ 
$f_{14}$ & 8.552 $\pm$ 0.002 & 0.10 & 0.027 &    5.0 &   3.2 &  4.06 $\pm$ 0.34
\enddata
\tablecomments{The formal errors on the photometric amplitudes are $\pm$
0.02~mmag; the 1-$\sigma$ errors on the radial velocity amplitudes are
$\pm$ 0.007~\kms. $f_{12}$ has been marked with a colon because it is not
fully resolved from $f_5$.}
\end{deluxetable}

\end{document}